\documentstyle[12pt,epsf]{article}
\begin{document}
\parskip=0pt
\parindent=123mm

{\hskip -10mm} TECHNION-PH-96-19/REV\par

\bigskip
\bigskip
\begin{center}
\parskip=10pt

{\huge    Bound on R-parity violating
couplings from nonleptonic B decay%$B^-\to K^-K^0 $
 }
\bigskip
\bigskip

{\large   Cai-Dian L\"u and Da-Xin Zhang }

 Physics Department, Technion- Israel Institute of Technology,
 Haifa 32000, Israel

\end{center}

\parindent=23pt
\bigskip
\begin{abstract}
The  effective Hamiltonian for
the R-parity violating couplings induced 
nonleptonic decay $B^-\to K^-K^0$ is calculated
including leading log QCD corrections
running from $m_{\tilde q}$, the mass of the up-type squark, to $m_b$.
Experimental upper limit of this decay
is used to give the bound on the 
R-parity violating couplings $\lambda''$.
\end{abstract}

\newpage

\section{Introduction}

In the supersymmetric generalizations of the standard model,
there is a kind of couplings which has no standard model analogue.
These novel couplings, 
called the R-parity (or the matter parity) violating couplings \cite{1,1a},
are not  forbidden by some fundamental principles.
In a variety of supersymmetric models \cite{2},
due to  additional  horizontal symmetries which are broken
at some scale between the supersymmetry breaking and the
Planck scales, 
the R-parity violating couplings are bounded to be small \cite{3}.
However, these kinds of theoretical bounds are quite model
dependent. They depend not only on the choices of the
horizontal symmetries but also on the concrete power
counting rules enforced on the superfields \cite{3}.
Some  model dependent cosmological  constraints,
{\it e.g.} $\lambda''<<10^{-7}$ given by the requirement 
that at the grand unification scale baryogenesis does not
get washed out,
can be also evaded\cite{cos}.

Lacking  reliable predictions for these R-parity 
violating couplings,
it becomes important to determine  or
bound them from  experiments.
In the literature some  phenomenological  analyses
have been given based on a variety of experiments,
from proton decays \cite{4}
%the lepton-number violating decay $\mu\to e\gamma$
to the precise measurements of the electroweak interactions at LEP \cite{5}
 (see also \cite{6}).
Because these measurements are performed at different scales
and the QCD couplings between these scales are not small,
it is necessary to normalize the bounds at a same scale to extract
informations  at   high energy
scales ({\it e.g.} the grand unification  or the Planck scale).
This is essential to have a better understanding
on the origins of these couplings.
Note that in some newly performed renormalization group analyses \cite{7},
the running of the R-parity violating couplings has  
been carried out between $m_Z$ and the grand unification  scale.

Recently,  it has been pointed out that nonleptonic decays
of the B mesons, 
such as $B^-\to K^-K^0$ whose decay rate
in the standard model is quite small, 
can also provide
useful bound on the baryon-number nonconserved
R-parity violating couplings $\lambda''$s \cite{8}.
In the present work, 
we also concentrate on this process.
We will calculate the leading log QCD corrections
to the effective Hamiltonian,
sum up the large logs of the form $\alpha_s{\rm log}(m_{\tilde q}/m_b)$
(here ${\tilde q}$ is an up-type squark)
using the renormalization group equations.
We  perform the hadronic analyses using the
perturbative quantum chromodynamics (PQCD) method \cite{9,10}
and then use the current upper limit of the branching ratio
to bound $\lambda''$s.

\section{Effective Hamiltonian with Leading Log Corrections}

The superpotential for the R-parity violating couplings 
which do not conserve baryon-number is
\begin{equation}
W=\lambda''_{ijk}{\bar U_i}{\bar D_j}{\bar D_k},
\end{equation}
where $\bar U$ and $\bar D$ are right-handed superfields
and $i,j,k$ are generation indices.
These superfields are defined in the basis in which
the quarks are in their mass eigenstates.
%  We assume that the lightest sparticle is ${\tilde q}$ 
%  and we will return to this point later.
We consider only the  lightest charge-$\frac{2}{3}$
right-handed squark (denoted as ${\tilde q}$) and we will return to this point later.
Following the effective field theory description,
${\tilde q}$ is integrated out at the scale of $m_{\tilde q}$ and the results are two 
 4-quark operators \cite{1a}:
\begin{eqnarray}
O_1 &=& (\bar d_R \gamma_ \mu s_R)(\bar s_R \gamma^\mu b_R),\\
O_2 &=& (\bar s_R \gamma_\mu s_R)(\bar d_R \gamma^\mu b_R).
\end{eqnarray}
Then the effective Hamiltonian is written as
\begin{equation}
{\cal H} = \frac{\lambda''_{qbs} {\lambda''_{qds}}^*}{2m_{\tilde q}^2} 
\left(C_1(\mu) O_1 +C_2(\mu) O_2\right).
\end{equation}
The Wilson coefficients $C_1(\mu)$, $C_2(\mu)$ are calculated at the scale of  
${\tilde q}$ mass as
\begin{equation}
C_1(m_{\tilde q}) =1, ~~C_2(m_{\tilde q}) =-1.
\end{equation}

The leading log QCD corrections are characterized by the 
 anomalous dimensions for these two operators. 
Calculating the Feynman diagrams
in Fig.1, we get the following anomalous dimension matrix:
\begin{equation}
\gamma=\frac{g_s^2}{8\pi^2}\left(
\begin{array}{cc}
-1 & 3\\
3 & -1
\end{array} \right).
\end{equation}
By applying the QCD renormalization group equations
\begin{equation}
\mu \frac{d}{d\mu} C_i(\mu)=\displaystyle\sum_{j}(\gamma^{T})_{
ij}C_j(\mu),\label{ren}
\end{equation}
we can get the coefficients of the two operators at some low energy scale $\mu$
(here $\mu\sim m_b$)
from  those at the scale $\mu=m_{\tilde q}$.
In the case that   ${\tilde q}$ is lighter than the top quark,
\begin{eqnarray}
C_{1} (m_b) &=&  \frac{1}{2} (C_1 (m_{\tilde q}) +C_2(m_{\tilde q})) 
\xi ^{6/23} +\frac{1}{2} (C_1(m_{\tilde q})-C_2(m_{\tilde q})) \xi ^{-12/23} ,\\
C_{2} (m_b) &=& \frac{1}{2} (C_1(m_{\tilde q})+C_2(m_{\tilde q})) \xi ^{6/23} 
-\frac{1}{2} (C_1(m_{\tilde q})-C_2(m_{\tilde q})) \xi ^{-12/23},
\end{eqnarray}
where $\xi=\alpha_s(m_{\tilde q})/\alpha_s(m_b)$.
Otherwise threshold effects of the top quark should be accounted for by
two step running:
\begin{eqnarray}
C_{1} (m_t) &=&  \frac{1}{2} (C_1 (m_{\tilde q}) +C_2(m_{\tilde q})) 
\eta ^{6/21} +\frac{1}{2} (C_1(m_{\tilde q})-C_2(m_{\tilde q})) \eta ^{-12/21} ,\\
C_{2} (m_t) &=& \frac{1}{2} (C_1(m_{\tilde q})+C_2(m_{\tilde q})) \eta ^{6/21} 
-\frac{1}{2} (C_1(m_{\tilde q})-C_2(m_{\tilde q})) \eta ^{-12/21},
\end{eqnarray}
with $\eta=\alpha_s(m_{\tilde q})/\alpha_s(m_t)$, and
\begin{eqnarray}
C_{1} (m_b) &=&  \frac{1}{2} (C_1 (m_{t}) +C_2(m_{ t})) 
\eta_2 ^{6/23} +\frac{1}{2} (C_1(m_{ t})-C_2(m_{ t})) \eta_2 ^{-12/23} ,\\
C_{2} (m_b) &=& \frac{1}{2} (C_1(m_{ t})+C_2(m_{ t})) \eta_2 ^{6/23} 
-\frac{1}{2} (C_1(m_{ t})-C_2(m_{ t})) \eta_2 ^{-12/23},
\end{eqnarray}
where $\eta_2=\alpha_s(m_t)/\alpha_s(m_b)$.
Note that by taking $m_{\tilde q}=100, 300$ or $500$GeV,
the corrected Wilson coefficient $C_1$ at the scale of $m_b$
is 1.34, 1.45 or 1.50, which shows  large QCD corrections.

\section{PQCD Calculations of the Decay Rate}

Low energy hadronic transitions can be analyzed
in the framework of PQCD \cite{9,10}.
In the present case,
the operator $O_2$ which gives the non-factorizable contributions
has the same Wilson coefficient as that of $O_1$ (up to a minus sign).
Thus it is reasonable to use the formalism
of Simma and Wyler \cite{10} where no explicit use of the
factorization hypothesis is  needed.
Following this description \cite{10}
we take the interpolating field of the B meson as
\begin{equation}
\psi_B=\displaystyle\frac{1}{\sqrt{2}}\displaystyle\frac{I_c}{\sqrt{3}}
\phi_B(x)\gamma_5({\not\! p}-m_B),
\label{bwf}
\end{equation}
and those of the $K^-$ and $K^0$ mesons as
\begin{equation}
\psi_{K^-}=\displaystyle\frac{1}{\sqrt{2}}\displaystyle\frac{I_c}
{\sqrt{3}}\phi_{K^-}(y)\gamma_5 \not\!p_{K^-},
\label{kwf}
\end{equation}
\begin{equation}
\psi_{K^0}=\displaystyle\frac{1}{\sqrt{2}}\displaystyle\frac{I_c}
{\sqrt{3}}\phi_{K^0}(z)\gamma_5 \not\! p_{K^0}.
\label{kwf2}
\end{equation}
Here $I_c$ is  an identity in the color space. 
We have made the approximation $m_K= 0$.

The decay amplitudes can be calculated from Fig 2. 
With the insertions of the operator $O_1$, 
the calculations of the four diagrams give:
\begin{eqnarray}
{\cal A}_a^1&=&\int_{0}^{1} [{\rm d}x][{\rm d}y][{\rm d}z]
{\rm Tr}\left[
\displaystyle\frac{i}{\not\! p_b-m_b}\left(i\displaystyle\frac{\lambda^a}
{2}\gamma_\alpha g_s \right)
\psi_B \left(i\displaystyle\frac{\lambda^a}{2}\gamma^\alpha g_s\right)
\psi_{K^-} % \right.\\&&\left. 
 \gamma_\mu P_R \right] {\rm Tr} \left[\psi_{K^0} \gamma^\mu P_R \right]
 \displaystyle\frac{i}{l_g^2},\nonumber\\
	{\cal A}_b^1&=&\int_{0}^{1} [{\rm d}x][{\rm d}y][{\rm d}z]
{\rm Tr}\left[ \psi_B \left(i\displaystyle\frac{\lambda^a}{2}\gamma^\alpha 
g_s\right)\psi_{K^-} \gamma_\mu P_R  \right]
Tr\left[\psi_{K^0} \left(i
\displaystyle\frac{\lambda^a}{2}\gamma_\alpha g_s \right) 
\displaystyle\frac{i}{\not\! p_d}
 \gamma^\mu P_R \right]\displaystyle\frac{i}{l_g^2},\nonumber\\
	{\cal A}_c^1&=&\int_{0}^{1} [{\rm d}x][{\rm d}y][{\rm d}z]
{\rm Tr}\left[ \psi_B \left(i\displaystyle\frac{\lambda^a}{2}\gamma^\alpha
g_s\right) \psi_{K^-} \gamma_\mu P_R \right]
Tr\left[\displaystyle\frac{i}{\not\! p_{\bar s}}  
\left(i\displaystyle\frac{\lambda^a}{2}\gamma_\alpha g_s \right) \psi_{K^0}
 \gamma^\mu P_R \right]\displaystyle\frac{i}{l_g^2},\nonumber\\
	{\cal A}_d^1&=& \int_{0}^{1} [{\rm d}x][{\rm d}y][{\rm d}z]
{\rm Tr}\left[ \psi_B \left(i\displaystyle\frac{\lambda^a}
{2}\gamma^\alpha g_s\right) \psi_{K^-} \left(i\displaystyle\frac{\lambda^a}
{2}\gamma_\alpha g_s
 \right) \displaystyle\frac{i}{\not\! p_s}\gamma_\mu P_R \right] {\rm Tr}
\left[\psi_{K^0} \gamma^\mu P_R \right] \displaystyle\frac{i}{l_g^2},
\label{ampli1}
\end{eqnarray}
where $[ {\rm d}x]$, $[ {\rm d}y]$, and $[ {\rm d}z]$ denote 
$({\rm d}x_1{\rm d}x_2)$, $({\rm d}y_1{\rm d}y_2)$ and
$({\rm d}z_1{\rm d}z_2)$, respectively,
and
\begin{equation}
l_g=y_1p_--x_1p_B, ~~p_b=x_2p_B-l_g, ~~p_d=z_2p_0+l_g, ~~p_{\bar s}
=-z_1p_0-l_g, ~~p_s=y_2p_-+l_g,\label{prop}
\end{equation}
$p_-$ and $p_0$ being the momenta of the $K^-$ and  the $K^0$ mesons,
respectively.
 With the insertions of  the operator $O_2$, the results are:
\begin{eqnarray}
{\cal A}_a^2&=&\int_{0}^{1} [{\rm d}x][{\rm d}y][{\rm d}z]
{\rm Tr}\left[
\displaystyle\frac{i}{\not\! p_b-m_b}\left(i\displaystyle\frac{\lambda^a}
{2}\gamma_\alpha g_s \right)
\psi_B \left(i\displaystyle\frac{\lambda^a}{2}\gamma^\alpha g_s\right)
\psi_{K^-}  
 \gamma_\mu P_R \psi_{K^0} \gamma^\mu P_R \right] \frac{-i}{l_g^2},\nonumber\\
	{\cal A}_b^2&=&\int_{0}^{1} [{\rm d}x][{\rm d}y][{\rm d}z]
{\rm Tr}\left[ \psi_B \left(i\displaystyle\frac{\lambda^a}{2}\gamma^\alpha 
g_s\right)\psi_{K^-} \gamma_\mu P_R \psi_{K^0} \left(i
\displaystyle\frac{\lambda^a}{2}\gamma_\alpha g_s \right) 
\displaystyle\frac{i}{\not\! p_d}
 \gamma^\mu P_R \right]\displaystyle\frac{-i}{l_g^2},\nonumber\\
	{\cal A}_c^2&=&\int_{0}^{1} [{\rm d}x][{\rm d}y][{\rm d}z]
{\rm Tr}\left[ \psi_B \left(i\displaystyle\frac{\lambda^a}{2}\gamma^\alpha
g_s\right) \psi_{K^-} \gamma_\mu P_R \displaystyle\frac{i}{\not\! p_{\bar s}}  
\left(i\displaystyle\frac{\lambda^a}{2}\gamma_\alpha g_s \right) \psi_{K^0}
 \gamma^\mu P_R \right]\displaystyle\frac{-i}{l_g^2},\nonumber\\
	{\cal A}_d^2&=& \int_{0}^{1} [{\rm d}x][{\rm d}y][{\rm d}z]
{\rm Tr}\left[ \psi_B \left(i\displaystyle\frac{\lambda^a}{2}\gamma^\alpha
g_s\right) \psi_{K^-} \left(i\displaystyle\frac{\lambda^a}{2}\gamma_\alpha g_s
 \right) \displaystyle\frac{i}{\not\! p_s}\gamma_\mu P_R \psi_{K^0}
 \gamma^\mu P_R \right] \displaystyle\frac{-i}{l_g^2},
\label{ampli2}
\end{eqnarray}

Performing the trace operations in both the spinor and the color spaces, 
we find that the contributions of ${\cal A}_b^1$, ${\cal A}_c^1$ vanish
due to their color  structures.
The  amplitudes ${\cal A}_d^1$ and ${\cal A}_d^2$
are proportional to $m_K$ and thus vanish under the approximation
of $m_K= 0$. The amplitude ${\cal A}_a^1$ is the same as 
${\cal A}_a^2$ up to a color factor. 
Denoting ${\cal A}_a=C_1(m_b){\cal A}^1_a+C_2(m_b){\cal A}^2_a$, 
we get
\begin{eqnarray}
{\cal A}_a &=& -2\int_0^1dxdydz \frac{8g_s^2}{3\sqrt{6}} C_2(\mu)\phi_B(x)\phi_{K^-}(y)
\phi_{K^0}(z) \frac{2x_1-y_1-1}{x_1(2x_1-x_1^2-y_1)(y_1-x_1)},\\
{\cal A}_b&=& -\int_0^1dxdydz \frac{8g_s^2}{3\sqrt{6}} C_2(\mu)\phi_B(x)\phi_{K^-}(y)
\phi_{K^0}(z)\frac{1}{x_1(y_1-x_1)^2},\\
{\cal A}_c&=& \int_0^1dxdydz \frac{8g_s^2}{3\sqrt{6}}C_2(\mu) \phi_B(x)\phi_{K^-}(y)
\phi_{K^0}(z) \frac{2x_1-y_1-z_1}{x_1(x_1-z_1)(y_1-x_1)^2},\\
{\cal A}_d&=& 0.
\end{eqnarray}
Now the total decay amplitude reads
\begin{equation}
{\cal A}~\equiv\frac{\lambda''_{qbs} {\lambda''_{qds}}^*}{2m_{\tilde q}^2} 
({\cal A}_a+{\cal A}_b+{\cal A}_c+{\cal A}_d).
\end{equation}

The analyses given above are independent of the choice of the wave functions.
Below we will take the wave functions of the B and $K$ mesons  as \cite{9,10}
\begin{eqnarray}
\phi_{B}(x)&=&\frac{f_B}{2\sqrt{3}}\delta(x_1-\epsilon_B),\\
\phi_{K^-}(y)&=& \sqrt{3} f_{K} y_1 (1-y_1),\\
\phi_{K^0}(z)&=& \sqrt{3} f_{K} z_1 (1-z_1).
\end{eqnarray}
Now, after integrating over $[{\rm d}x]$, $[{\rm d}y]$ and $[{\rm d}z]$, 
we get the final amplitudes:
\begin{eqnarray}
&{\cal A}_a= -2\left (-\frac{1}{2}+2\ln\frac{1-2\epsilon_B}{2\epsilon_B}
-\ln\frac{1-\epsilon_B}{\epsilon_B}+i\pi \right ) G,&\nonumber\\
&{\cal A}_b + {\cal A}_c =\frac{3}{2}G,&
\end{eqnarray}
where
\begin{equation}
G=\frac{8 \pi \alpha_s(\mu)}{9\sqrt{2}\epsilon_B}C_2 f_Bf_{K}^2 .
\end{equation}
The decay width of $B^-\to K^-K^0$ is then
\begin{equation}
\Gamma =\frac{|\lambda''_{qbs} {\lambda''_{qds}}^*|^2 |{\cal A}_a +{\cal A}_b 
+ {\cal A}_c|^2}{64\pi m_B m_{\tilde q}^4}.
\end{equation}
Taking the $B^-$ meson lifetime as $1.54\times 10^{-12}s$, the branching ratio
of this decay can be easily found.

\section{Result and Conclusion}

The present experimental upper bound for the branching ratio 
of the decay $B^-\to K^-K^0$ is $5\times 10^{-5}$ \cite{exp}.
This gives the constraints for the R-parity violated coupling 
$|\lambda''_{qbs} {\lambda''_{qds}}^*|$
 depending on the mass of ${\tilde q}$  (see equations (8) - (13) and (28) - (30)). 
As we have stated previously,
we have considered only the lightest  up-type squark.
When the masses of  the supersymmetric partners of the up, the charm 
and the top quarks are comparable with one another,
none of their contributions can be neglected.
In this case,  there exists the possibility of cancellations between
the contributions from these up-type squarks,
and what we get is the bound on 
$|\sum_{i=u,c,t}C_1^i\lambda''_{ibs} {\lambda''_{ids}}^*/m_{\tilde i}^2|$.

The  excluded region in the parameter space is shown in Fig.3.
For the numerical estimations, we have used parameters as
$\alpha_s(M_Z)=0.117$, the hadronic scale $\mu=1$GeV, and the decay 
constants are taken to be $f_B=132$MeV and $f_{K}=113$MeV. The value of 
$\epsilon_B$ characterizes most of the uncertainties in the PQCD calculations.
Note that if we take $\epsilon_B=0.05$, 
the bound without QCD corrections coincides with what is
given in \cite{8}. 
Here we also give the bound by using 
$\epsilon_B=0.065$ which is 
favored by the rare decay of $B\to K^* \gamma$ \cite{lz}.

As a conclusion, 
the effects of QCD corrections to the effective Hamiltonian induced 
by the R-parity violating couplings  are significant. 
The bound on a particular combination of
these couplings has been upgraded by including the
QCD corrections. 

\section*{Acknowledgement}

We thank B. Blok and C. Liu for helpful discussions.
The research of D.-X. Z. is supported in part by Grant 5421-3-96
from the Ministry of Science and the Arts of Israel.

\bigskip
\noindent
{\large Figure Captions:}

Fig. 1 Diagrams in the calculation of the anomalous dimensions of the operators. 
The black blob represents the effective operator $O_1$ or $O_2$.

Fig. 1 Diagrams contributing to $ B^-\to K^- K^0 $ in PQCD.
The black blob 
represents the effective operator $O_1$ or $O_2$.

Fig.3 Bounds on $|\lambda''_{qbs} {\lambda''_{qds}}^*|$ from the experimental 
upper limit of $ B^-\to K^- K^0 $ as a function of ${\tilde q}$ mass.
The upper two lines represent results without QCD corrections, the dash-dotted 
line with $\epsilon_B=0.05$, the dotted line with $\epsilon_B=0.065$.
The lower two lines represent results with QCD corrections, with the 
solid one for $\epsilon_B =0.05$ and the dashed one for $\epsilon_B=0.065$. 

\begin{figure}
\centerline{\epsffile{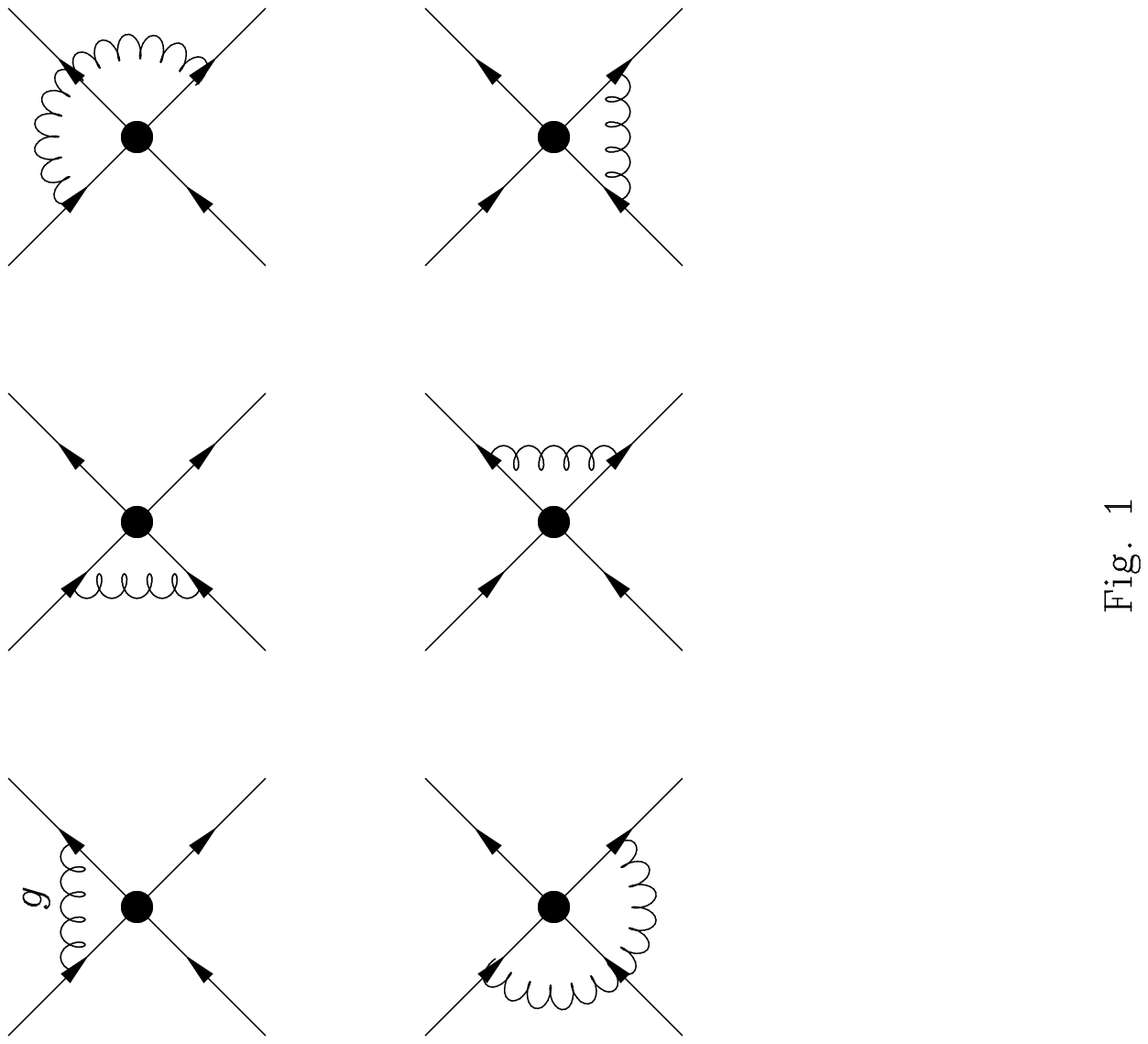}}
\end{figure}

\begin{figure}
\centerline{\epsffile{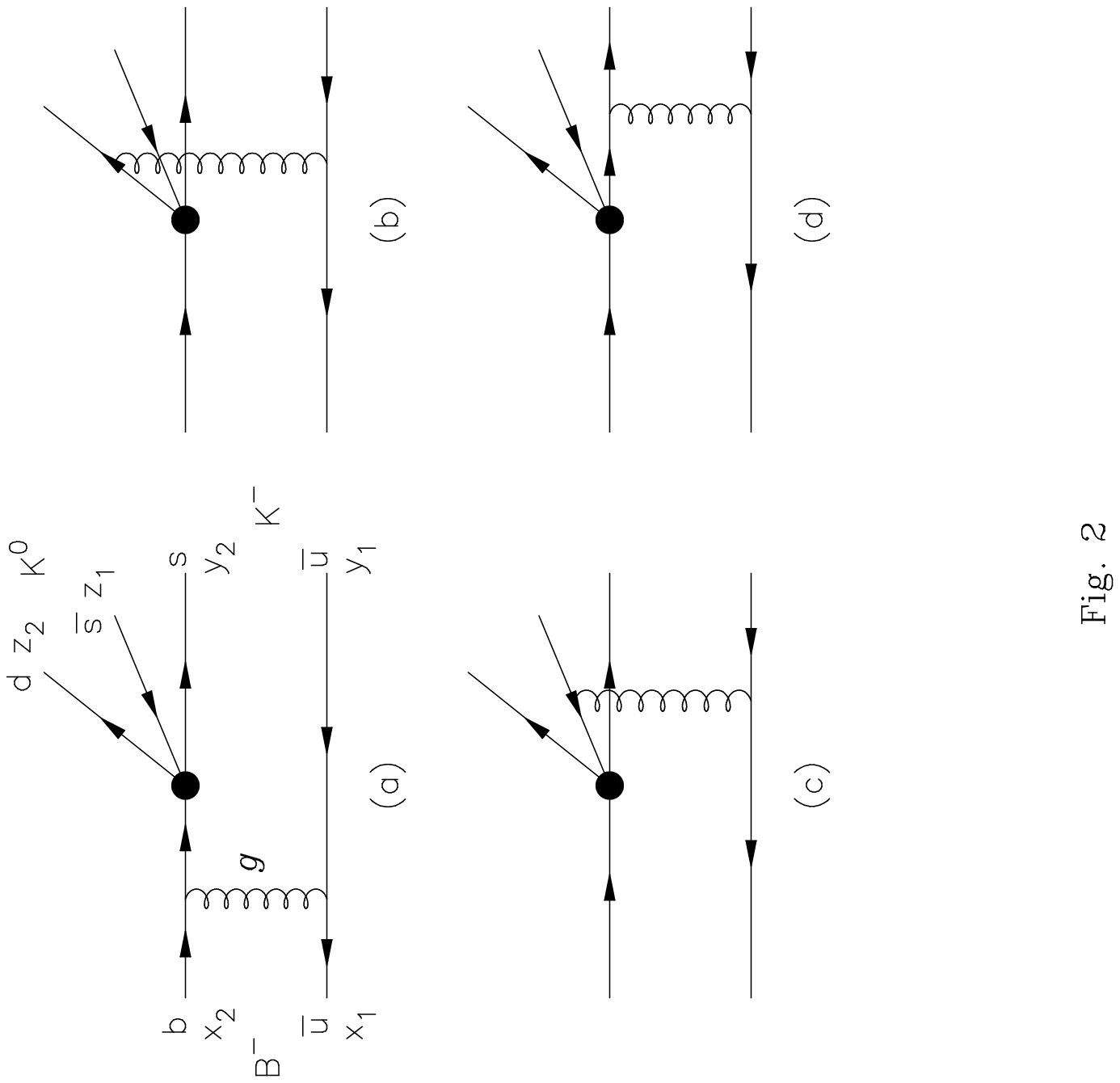}}
\end{figure}

\begin{figure}
\centerline{\epsffile{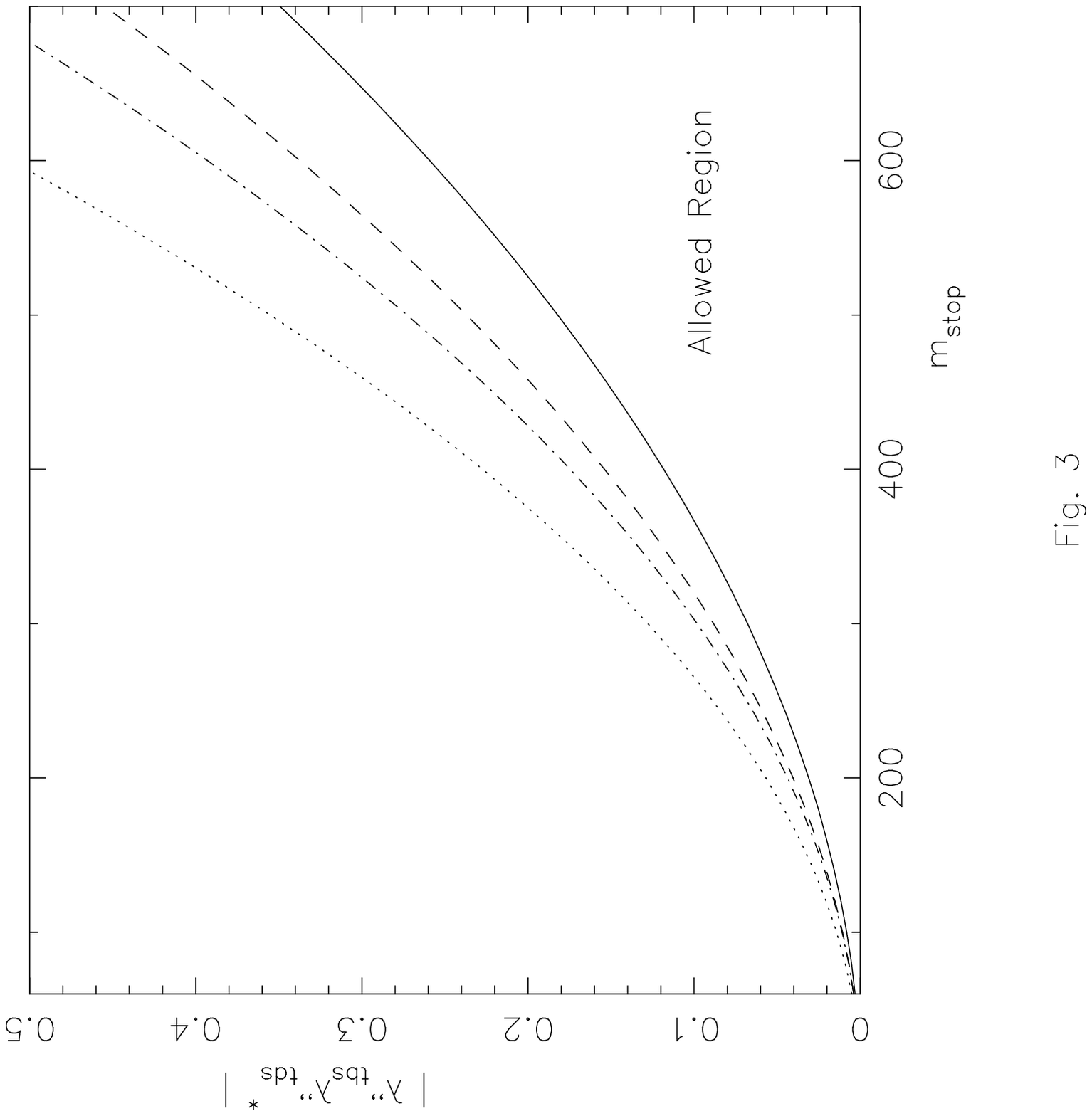}}
\end{figure}


\newpage

\begin{thebibliography}{99}
\bibitem{1}
J. Ellis, G. Gelmini, C. Jarlskog, G.G. Ross and J.W.F. Valle,
Phys. Lett.  B150, 142 (1985);
G.G. Ross and J.W.F. Valle, Phys. Lett.  B151, 375 (1985);
L Hall and M. Suzuki, Nucl. Phys. B231, 419 (1984);
I-H. Lee, Nucl. Phys. B246, 120 (1984);
S.  Dawson,  Nucl .Phys. B261, 297 (1985).
For a review, see R. Vissani, talk given at SUSY 96, preprint hep-ph/9607423.

\bibitem{1a}
 R. Barbieri and  A. Masiero, Nucl. Phys. B267, 679 (1986). 

\bibitem{2}
Y. Nir and N.  Seiberg, Phys. Lett. B309, 337 (1993);
M. Leurer, Y. Nir and N.  Seiberg, Nucl. Phys. B420, 468 (1994);
M. Dine, R. Leigh and A. Kagan, Phys. Rev. D48, 4269 (1993);
C.  D. Carone, L.  J. Hall and H.  Murayama,  Phys. Rev. D54, 2328 (1996).

\bibitem{3}
V. Ben-Hamo and Y. Nir, Phys. Lett.  B339, 77 (1994).

\bibitem{cos}
H. Dreiner, G.G. Ross,  Nucl. Phys. B410, 188 (1993).

\bibitem{4}
A. Yu. Smirnov and F. Vissani, Nucl. Phys. B460, 37 (1996);
preprint IC-96-16,  hep-ph/9601387.

\bibitem{5}
G. Bhattacharyya, D. Choudhury and  K. Sridhar,  Phys. Lett. B355, 193 (1995);
P. H. Chankowski,  D. Choudhury and S. Pokorski, preprint
SCIPP-96-27, hep-ph/9606415. 

\bibitem{6}
Other bounds can be found in {\it e.g.},
K. Enqvist, A. Masiero and A. Riotto, Nucl. Phys. B373, 95 (1992);
 J.L. Goity and M. Sher, Phys. Lett. B346, 69 (1995);
M. Chaichian and K. Huitu, preprint  HU-SEFT-R-1996-09,  hep-ph/9603412;
H. Nunokawa, A. Rossi and J.W.F. Valle, preprint FTUV-96-33, hep-ph/9606445.

\bibitem{7}
H. Dreiner and H. Pois, preprint  ETH-TH-95-30, hep-ph/9511444;
V. Barger,  M.S. Berger,  R.J.N. Phillips and T. Wohrmann,
Phys. Rev. D53, 6407 (1996);
B. de Carlos and  P.L. White, preprint SUSX-TH-96-003, hep-ph9602381.

\bibitem{8}
C.E. Carlson, P. Roy and M. Sher, Phys. Lett. B357, 99 (1995).

\bibitem{9}
 A. Szczepaniak, E. M. Henley and S. J. Brodsky, Phys. Lett. B243, 287 (1990);
C.E. Carlson and   J. Milana, Phys. Rev. D49, 5908 (1994); Phys. Lett. B301,
237 (1993).


\bibitem{10}
H. Simma and D. Wyler, Phys. Lett.  B272, 395 (1991).

\bibitem{exp}
D.M. Asner {\it et al.} (CLEO Collaboration), Phys. Rev. D53, 1039 (1996).
See also \cite{8} for a discussion.

\bibitem{lz} C.D.  L\"{u}
and D.X. Zhang, preprint TECHNION-PH-96-13.

\end{thebibliography}
\end{document}